 \documentclass[prl,aps,twocolumn,showpacs]{revtex4}
\usepackage[dvips]{graphicx}
\usepackage{dcolumn}
\newcommand{\be}{\begin{equation}}
\newcommand{\ee}{\end{equation}}
\newcommand{\bea}{\begin{eqnarray}}
\newcommand{\eea}{\end{eqnarray}}
\newcommand{\nn}{ \nonumber}
\newcommand{\ds}{\displaystyle}
\begin{document}
\topmargin=-20mm
%   \Large

\title {Local Geometry of the Fermi Surface and Quantum Oscillations in the Linear Response of Metals }

\author{ Natalya A. Zimbovskaya}

\affiliation{Department of Physics and Asronomy, St. Cloud State University,  St. CLoud, MN 56301, USA, and Department of Physics and Electronics, University of Puerto Rico at Humacao, PR 00791} 

 \begin{abstract} %1
 In this paper we present a theoretical analysis of the 
effect of anomalies  of the Fermi surface (FS) curvature on  
oscillations in the electron density of states (DOS) in 
strong magnetic fields. It is shown that the oscillations 
could  be significantly weakened when the FS curvature  
takes on zero or extremely large values or diverges in the 
vicinities of some extremal cross-sections. This 
leads to a characteristic dependence of the oscillations 
magnitude on the direction of the magnetic field which is 
studied. The results of these general studies are employed 
to analyze the effects of the FS curvature anomalies on the 
quantum oscillations in the velocity of sound waves 
travelling in metals. 
  \end{abstract}

\pacs{71.18.+y, 71.20-b, 72.55+s}
\date{\today} 
\maketitle 

\section{1. Introduction}

Experimental data concerning quantum oscillations in various 
observables in metals under strong magnetic fields were 
repeatedly used in studies of their electron characteristics 
\cite{1}.
It is known that the electron response of a metal to an 
external disturbance depends on the geometry of the FS. 
The Fermi surfaces of  metals are mostly  complex in 
shape and this can essentially influence observables. 
At present those 
phenomena which are determined by the main geometric 
characteristics of the FS's, i.e. their connectivity, are 
well studied. However those effects which appear due to 
some fine geometrical features of the FSs such as nearly cylindrical segments or points of flattening have not been investigated in detail so 
far. Meanwhile these elements in the FS geometry
can noticeably affect the electron response of 
metal.

When the FS includes points and/or lines where its curvature turns zero, this it leads to an enhancement of the contribution from the neighborhood of these points/lines to the electron density of states (DOS) on the 
FS. Usually this enhanced contribution is small compared to 
the main term in the DOS which originates from the 
remaining major part of the FS. Therefore it cannot produce 
noticeable changes in the response of the metal  when all 
segments of the FS contribute to the response functions 
essentially equally. 
However, some effects (including those originating from 
quantum oscillations in the electron DOS) are determined 
with  small ``effective" segments of the FS. The 
contribution to the DOS 
from the vicinities of the points and lines of zero curvature can be 
congruent to the contribution of an  ``effective'' 
segment.  In other words, when the curvature of the surface
becomes zero at some part on an ``effective'' segment of 
the FS, it can give a sensible enhancement of efficient 
electrons and, in consequence, a pronounced change in the
response of the metal to the disturbance.

It has been shown that when the FS includes points or lines 
where at least one of the principal radii  of curvature 
becomes infinitely large, then
changes may be observed in the frequency and temperature 
dependences of sound  dispersion and  absorption \cite{2,3,4,5,6}, and in the frequency dependence of the 
surface impedance of a conventional metal under anomalous skin effect \cite{7,8}. Likewise, the flattening points at the FS may give rise to some anomalies in the magnetoacoustic response of a two-dimensional electron gas, as shown in the papers \cite{9,10}.

In the present work we concentrate on quantum oscillations in observables in metals which originate from the DOS oscillations in strong magnetic fields. The oscillating term in the DOS is formed by the neighborhoods of the FS cross-sections with extremal areas. Usually, the latter are narrow strips on the FS. Therefore the FS curvature anomalies at these strips may significantly affect their contributions to the DOS quantum oscillations. Consequently, the effect of the FS local geometry can be revealed in oscillations of various observables in metalls, such as magnetization, magnetic susceptibility, resistivity, sound velocity and attenuation.  
Qualitative anomalies of the de-Haas-van Alphen oscillations associated with cylindrical pieces of the FSs were considered before in Refs. \cite{11,12,13}. 
Also, it was shown that nearly cylindrical pieces inserted in the FS could cause noticeable softening of some acoustic modes near the peaks of the DOS quantum oscillations (see Refs. \cite{14,15}).

Along with the points and lines of zero curvature there can 
exist points and lines, where the FS curvature diverges.
A  simple illustration  is a kink line on a  
surface where the electron velocity is discontinuous.
 If the FS inserts such lines, then the effective 
belts could be missed for appropriate directions of 
propagation of electromagnetic or sound waves, which brings 
anomalies in high-frequency properties of the metal. It is 
shown  \cite{16,17} that under these conditions 
electromagnetic waves of some special kind can propagate in
metals. The lines of a singular (infinite) curvature can be arranged on edges of narrow lenses or needle-shaped cavities, which are elements of the FSs of some metals. However, it could 
happen that the FS curvature diverges at some points where the velocity of electrons varies continuously  
as discussed below.   

A characteristic feature of quantum oscillations attributed 
to the extremal cross sections of anomalous curvature is a 
clearly manifested dependence of their amplitude on the 
direction of an external magnetic field $\bf B$. The effect 
could be revealed only for certain directions, 
when the effective cross-section runs along that part 
of the FS where the curvature  turns zero or 
diverges. When the magnetic field is tilted away from such 
direction by an angle $ \Phi $ the effective cross section slips from the ``anomalous" segment of the FS. This brings sensible changes in the oscillations magnitude.  This is typical for all effects arising due to local peculiarities in the FS geometry. For instance, angular dependence of the amplitudes of magnetoacoustic commensurability oscillations in a two-dimensional electron system was discussed in Ref \cite{10}. In the present work we analyze  angular dependencies 
of quantum oscillations magnitudes for different kinds of 
local anomalies of the FS curvature. 
The results could be employed to analyze manifestations of the FS local geometry in quantum oscillations of various characteristics in metals. As an example, we briefly consider the effect of the FS curvature anomalies on quantum oscillations in the velocity of an ultrasonic wave propagating in a metal. To avoid further prolixities in this work, other applications are not included. These applications are to be discussed elsewhere.

\section{2. The model}

The concept of a zero curvature line (nearly cylindrical cross-section) on the FS does not require further explanation. On the contrary, the concept of the line where the FS curvature diverges needs clarification. Before introducing the FS model used in the following analysis, we present a simple illustration to clear this concept.

  It is known that FSs of cadmium and zinc (both metals have a hcp crystalline structure) include an axially symmetric electron lens. The lens is located at the center of third Brillouin zone, with its axis running along the [0001] direction. We can roughly reproduce this lens using the Harrison method of construction of free electron FSs. Assuming that ``z" axis of the coordinate system is directed along the axis of symmetry, we arrive at the following energy-momentum 
relation for electrons associated with the lens:
  \be %f1
 E({\bf p}) = \frac{{\bf p}^2_\perp}{2m} + 
\frac{p^2_m}{2m} \left (sgn (x) \cdot x + \frac{hG}{2 p_m} 
\right )^2 .
    \ee
           Here, $\bf p_\perp $ and $ p_z$ are the electron 
quasimomentum components across and along the symmetry axis, 
respectively: $ m $ is the electron effective mass; $ x = 
p_z / p_m $ where $ p_m$ is the maximum value of the 
quasimomentum along the "z" axis; and $ {\bf G } = (0,0,G) 
$ is the  corresponding reciprocal lattice vector.

The lens described with (1) has a kink line running 
along its edge. However, this curvature anomaly appears 
owing to our free electron approximation. In actual metals 
we can expect it to be smoothed by means of crystalline 
fields. To better reproduce the shape of the lens near the 
edge we start from the standard asymptotic for the sign 
function which is valid for $ |x| < 1, $ namely:
  \be %f2
  sgn (x) = \lim_{l \to \infty} f (x,l) \equiv \lim_{l \to \infty} |x|^{1/l + 1}/x.
  \ee
      To proceed we use the approximation $ sgn (x) \approx 
f (x,l). $ Substituting this approximation into (1) we see 
that now the velocity of the electrons belonging to the lens 
varies continuously, and its longitudinal component reduces 
to zero at the central cross-section $ (p_z = 0 )$. Keeping 
only the greatest terms, we can write the following 
expression for the cross-sectional area near the central 
cross-section:
  \be %f3
 A(x) = A_{ex}(1 - b^2 |x|^{2k} ).
  \ee
  Here, $A_{ex} $ is the central cross-sectional area, 
$b^2$ is a dimensionless positive constant, and $ k = 
\frac{1}{2} (1 + 1/l). $ The lens curvature near the 
central cross-section could be approximated as follows:
    \be %f4
  K(x) \approx  - \frac{1}{2 p_m^2 A_{ex}} 
\frac{d^2 A}{dx^2} \sim \frac{|x|^{2k-2}}{p_m^2}.
  \ee
 Since $ k < 1, $ the curvature tends to infinity when 
$ p_z $ tends to zero although these is no break in the 
longitudinal velocity at $ p_z = 0 $, as shown in the Fig. 1. This means that the central cross-section of our FS is not a kink line, and the effective strip still exists here.  However, there could be a considerable decrease in the number of electrons associated with the vicinity of this cross-section. It can influence some properties of the metal including quantum oscillations. 
 Actually, there exists a piece of experimental evidence for 
the curvature anomaly at the edge of the electron lens in 
cadmium. A resonance feature at the cyclotron frequency was 
observed in the surface impedance derivative when the 
magnetic field was directed along the [0001] axis \cite{18}. 
This resonance originates from the contribution of the 
electrons associated with the edge of the lens, and it could 
be explained assuming that the lens curvature diverges 
there  \cite{19}.

\begin{figure}[t]  % fig.1a,b,c,d
\begin{center}
\includegraphics[width=8.6cm,height=8cm]{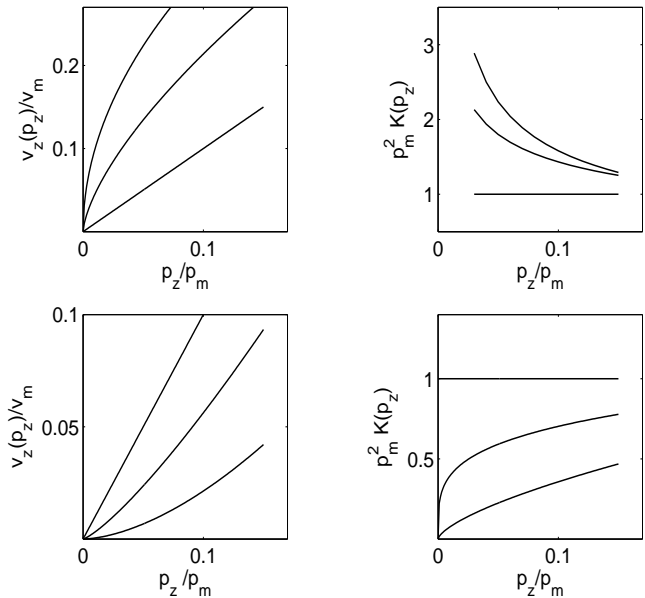}
%[width=5.0cm,height=6.6cm]{M1.eps}%{g2.eps}
\caption{
Curvature anomalies at axially symmetric Fermi surface. Top panels: Plots of $v_z $ versus $ p_z $ (left panel) and $ K $ versus $p_z $ (right panel) in the vicinity of the extremal cross-section at $ p_z = 0 $ where the FS curvature $ K $ diverges. The curves are plotted according to Eqs.(4) and (5) for $ m_\perp = m_{||}; \ k = 3/4, 5/6, 1 $ from the top to the bottom. Bottom panels: Plots of $ v_z $ versus $ p_z $ (left panel) and $ K $ versus $ p_z $ (right panel) in the vicinity of the nearly cylindrical cross-section at $ p_z = 0. $ The curves are plotted for $  m_\perp = m_{||}; \ k = 1, 9/8, 4/3 $ from the top to the bottom. 
}  
\label{rateI}
\end{center}
\end{figure}

In further calculations we use a simple 
model of a closed simply connected and axially symmetric FS 
which has a unique extremal cross section for any 
direction of the external magnetic field. 
 Such FS could be described with the energy-momentum relation:
  \be %f5
 E {\bf (p)} = \frac{{\bf p}_\perp^2}{2 m_\perp} + \frac{p_m^2}{2m_{||}} |x|^{2k} .
   \ee
 When $ k = 1, $ Eq.(5) corresponds to the ellipsoidal FS. In this case $ m_\perp, m_{||} $ are the principal values of the effective mass tensor. As previously, the cross-sectional area of the FS near its central cross-section is described with the expression (3), where $ b^2 = 1. $ It follows from (4) that at $ \Phi = 0  $ (the magnetic field is directed along the symmetry axis), the chosen FS is nearly cylindrical in the vicinity of the central cross-section when $ k > 1. $ However, for $ 1/2 < k < 1 $ the FS curvature diverges here. Suppose that the magnetic field $ \bf B $ deviates from the symmetry axis by the angle $ \Phi. $ Then the expression for the cross-sectional area changes. After some straightforward calculations we arrive at the result:
 \bea% f6,7  
 A(x) &=&A_{ex}(1 - a^2x^2 - b^2x^{2k}); \nn\\
  \qquad k &>& 1/2. 
             \eea %           
   Here,  $ a^2 , b^2 $ and $ A_{ex} $ are functions of the angle $\Phi $ between the magnetic field  and the FS 
symmetry axis,  $ a^2(0) = 0, $ and it increases whereas $\Phi $ enhances. When  the angle takes on a certain value $ \Phi_0 , \ a^2 (\Phi)$ becomes  greater than  $ b^2 (\Phi), $ and 
stays likewise when $ \Phi $ further increases.

\section
{3. The effect of the Fermi surface curvature anomalies on the quantum oscillations in the electronic density of states}

 For FSs with a single extremal cross-section the electron DOS is given by:
  % f7,10,5
 \be
N_\zeta \equiv - \sum_\nu \frac{d f_\nu}{\partial E_\nu}
= g(1 + \Delta).
  \ee
 Here, $ f_\nu $ is the Fermi distribution function for 
quasiparticles with energies $ E_\nu; \ g $ is the value of 
the electron DOS in the absence of the magnetic field, and 
the oscillating function $\Delta$ is described by 
the expression
  % f8,11,6
  \be 
\Delta = \sum\limits_{r=1}^\infty (-1)^r \psi_r(\theta)  
\cos\left[\pi r\frac{m_{\perp }}{m} \right]  
%$$$$ \times
 \int_0^1
\cos\left[\frac{rcA(x)}{\hbar|e|B}\right]dx,
                        \ee
 where $\ds  \psi (\theta)= {r \theta}/{\sinh r \theta}; 
\ \theta = {2 \pi^2 T}/{\hbar \Omega}; \, \ \hbar 
\Omega $ is the  cyclotron quantum; $T$ is the temperature 
expressed in energy units. 
 
Substituting (6) into (8), we obtain
   % f9,12
   \bea
\Delta &=& \sum\limits_{r=1}^\infty (-1)^r \psi_r(\theta)
\cos\left(\pi r\frac{m_\perp}{m}\right)
 \nn \\  & \times &
 \big[\cos(\pi r\gamma^2)U_r(\Phi) +
\sin(\pi r\gamma^2)W_r(\Phi) \big].
  \eea     
 Here,
      \bea % f10,13   
U_r(\Phi) &=& \int_0^1\cos \left [\pi r\gamma^2
(a^2x^2 + b^2 x^{2k})\right]dx,                               
   \\ %\ee
    % f11,14
    %\be      
 W_r(\Phi) &=& \int_0^1\sin \left [\pi r\gamma^2
(a^2x^2 + b^2 x^{2k}) \right]dx .
    \eea                                    
 In further analysis we assume that the number of 
Landau levels under the FS is large, and the 
parameter $\ds \gamma^2 = 2 \zeta/\hbar \Omega \ (\zeta $ is 
the chemical potential of electrons) takes values much 
larger than unity. So, we can derive asymptotic expressions 
for the integrals (10), (11) using the stationary phase method.

The principal terms in the expansions of these integrals 
in powers of the small parameter $\gamma^{-1} $ are 
easily  obtained in two limiting cases: $ a^2 \ll b^2$ and 
$ a^2 \gg b^2. $ The first inequality is satisfied for small 
values of the angle between the magnetic field and symmetry
axis, when the extremal cross-section of the FS 
nearly coincides with the line of anomalous curvature.
The asymptotic expression for the oscillating function 
$ \Delta $  has the form
     % f12,15,11
  \be
\Delta = \frac{\eta_k}{\gamma^{1/k}}
\sum\limits_{r=1}^\infty \frac{(-1)^r}{r^{1/2k}}
\psi_r(\theta)
\cos\left[\frac{rcA_{ex}}{\hbar|e|B} - \frac{\pi}{4k}\right]
\cos\left[\pi r\frac{m_\perp}{m}\right],
       \ee                           
     where
 $$
\eta_k = \frac{\Gamma(1/2k)}{2k(b\sqrt\pi)^{1/k}} \frac{m_\perp p_m}{\pi^2 \hbar^3 g}.
      $$
\begin{figure}[t]
\begin{center}
\includegraphics[width=8.4cm,height=4.2cm]{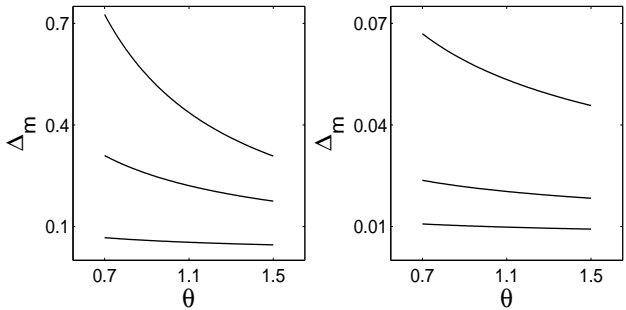}
%[width=5.0cm,height=6.6cm]{M1.eps}%{g2.eps}
\caption{
Effect of the FS curvature on the temperature dependence of quantum oscillation amplitudes at moderately low temperatures $ (\theta \sim 1) .$ Left panel: nearly cylindrical cross-sections; $ k =4,2,1 $ from the top to the bottom. Right panel: cross-sections where the FS curvature diverges; $ k = 1,3/4,5/8 $ from the top to the bottom. Curves are plotted assuming $ m_{||} = m_\perp,\ \pi \gamma^2 = 10^3. $
}  
\label{rateI}
\end{center}
\end{figure}                                          

In the opposite limit $ a^2 \gg b^2,$ which corresponds to 
 sufficiently large values of the  angle $ \Phi $, 
the principal term in the expansion of $ \Delta $ in powers 
of $\gamma^{-1} $ equals:
  % f13,16,12
 \be
\Delta = \frac{1}{\gamma}
\sum\limits_{r=1}^\infty
\frac{(-1)^r}{\sqrt r} \psi_r(\theta)
\cos\left[\frac{rcA_{ex}}{\hbar|e|B} - \frac{\pi}{4}\right]
\cos\left[\pi r\frac{m_\perp}{m}\right].
         \ee                        
 So, in this case, the function $\Delta$ describes ordinary
oscillations of the electron density of states in a 
quantizing magnetic field. 

It is well known that amplitudes of quantum oscillations depend on temperature. Ordinary quantum oscillations of the electron DOS described by the Eq. (13) have the amplitude of the order of $ \gamma^{-1/2} \theta^{-1/2}. $ The amplitude of the oscillations related to the ``anomalous" cross-section given by (12) may be estimated as $ \gamma^{-1/k} \theta^{(1-2k)/2k}. $ Comparing these estimates, we see that the amplitude of oscillations associated with the extremal cros-section of zero/diverging curvature significantly differs from that of the usual quantum oscillations. The contribution from a nearly cylindrical cross-section can considerably exceed in amplitude contributions from usual cross-sections.
As well, the amplitude of DOS oscillations originating from a cross-section where the FS curvature diverges is smaller than that of the ordinary oscillations.  Also, the fine geometrical  structure of the FS near the extremal cross-sections may bring changes in the temperature dependence of the oscillations amplitude. As shown in the Fig. 2, the closer is the vicinity of an extremal cross-section to a cylinder, the more pronounced is the decrease in the oscillation amplitude while the temperature rises. Oscillations associated with a cross-section where the FS curvature diverges, exhibit weaker temperature dependence compared to  ordinary quantum oscillations.

Now we analyze the amplitude of oscillations within the intermediate range where $ a^2$ and $ b^2 $ have the same 
order of magnitude. We present the integrals (10) and (11) 
as  expansions in powers of the parameter $ w \;(w = \pi 
r \gamma^2a^2/(\pi r\gamma^2b^2)^{1/k}).$ In particular, the 
expression for $ U_r $ in the case $ w < 1 $ has the form
   \bea %f14,17,13
U_r(\omega) &=& \frac{(\pi r\gamma^2b^2)^{-1/2k}}{2k}
\sum\limits_{n=0}^\infty \frac{(-1)^n}{n!} w^n
  \nn \\  &\times &
\Gamma\left(\frac{2n + 1}{2k} \right)
\cos\left[\pi\frac{n(1 - k)}{2k} + \frac{\pi}{4k}\right],
        \eea                                         
 while for $ w > 1$ we arrive at the following expansion:
   \bea  % f15,18,14
U_r(\omega)&=& \frac{1}{2\pi a\gamma\sqrt r}
\sum\limits_{n = 0}^\infty \frac{(-1)^n}{n!}w^{-nk}
   \nn\\  &\times&
\Gamma\left( kn +\frac{1}{2}\right)
\cos\left[\pi\frac{n(k - 1)}{2} + \frac{\pi}{4}\right].
             \eea                         
Expansions for $ W_r $ could be  obtained from (14) and 
(15) by replacing cosines with sines of the same arguments.

   The results of  computation of the oscillation 
amplitude  $ Y \ (Y = \sqrt{U^2 + W^2})$ as a function 
of $ w, $ are presented in the Fig. 2. 
Calculations were carried out assuming  that $ \theta 
\approx 1.$ Accordingly, only the first term was kept
 in the sum over $ r $ included in (12).  The dependence of $ Y $ on $ w $ calculated for $ k = 2$ and $ k = 4$ is shown in the left panel. In this case the central  cross section  of the FS for $\Phi = 0 $ is quasi-cylindrical.  The oscillations amplitude takes on the 
largest value when $ \Phi = 0 $ and decreases monotonically 
as $ w $ increases. Both curves are asymptotically 
approaching the straight line $Y = 1/\gamma $ when $ w \gg 1. 
$ As expected, the amplitude variation is 
much better pronounced for the curve 2. This curve is 
plotted for the larger value 
of the parameter $ k $ when the FS is closer to a 
cylinder near  its central cross-section.
In the right panel the behavior of the ampitude of quantum 
oscillations of the DOS upon a variation of the angle $\Phi$ is 
shown under the assumption that the FS curvature diverges
at the central  cross-section at $\Phi = 0.$ The curves presented here are plotted for $k = 3/4$ and $k = 5/8.$
The second value of $k$ corresponds to a stronger anomaly in 
the FS curvature at the extremal cross-section. For 
$k = 5/8,$ the oscillation amplitude increases monotonically 
with $w $ and tends to the value $1/\gamma.$ For a 
less pronounced anomaly in the curvature 
$(k = 3/4),$ the dependence of $Y$
on $w$ is nonmonotonic. However, in this case  the 
decrease in the value of $Y$ for small values of $ w $ is 
replaced by its increase when $ w $  takes on greater values.

\begin{figure}[t]
\begin{center}
\includegraphics[width=7cm,height=4.8cm]{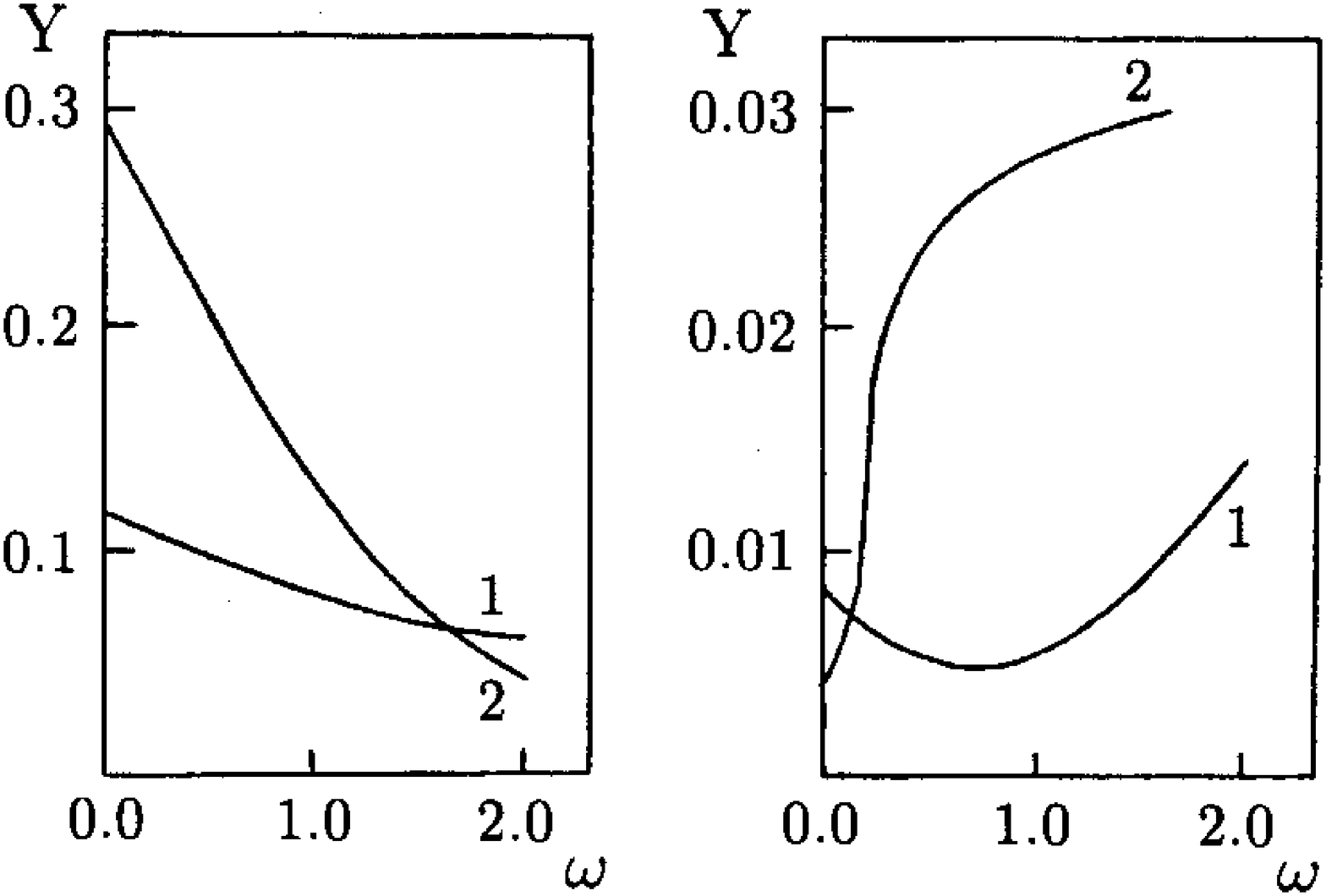}%{m2_5.eps}
%[width=5.0cm,height=6.6cm]{M1.eps}%{g2.eps}
\caption{
The amplitude of the electron DOS quantum oscillations as a function of the magnetic field orientation. Left panel: $k = 2$ (curve 1), $ k= 4$ (curve 2). Right panel: $ k = 3/4$ (curve 1), $k = 5/8$ (curve 2). The curves are plotted for $ \pi \gamma^2 = 10^3. $
}  
\label{rateI}
\end{center}
\end{figure}

  Now, we remark that two parameters characterizing the FS local geometry were used in the present work, and they both could be easily evaluated provided that we possess relevant experimental data. The angle $ \Phi_0 $ describing the width of the FS strip where the curvature anomalies are revealed is a directly measurable quantity, and the shape parameter $ k $ introduced in the Eqs. (3) and (5) could be estimated basing on the variation in the oscillations magnitude between its maximum/minimum $ (\Phi = 0) $ and plateau $ (\Phi \gg \Phi_0) $ values. The magnitudes of oscillations described by Eqs. (12),(13) have the order of $\gamma^{-1/k} \theta^{(1/2k)-1}, $ and $\gamma^{-1} \theta^{-1/2},$ respectively. So, the ratio of maximum/minimum and plateau magnitudes $ \rho $ takes on value: $\rho \sim \gamma^{1 - 1/k} \theta^{(1-k)/2k} $ which gives the following estimate for the shape parameter: $  k \approx \big (\frac{1}{2} \ln \theta - \ln \gamma \big )\big /\big (\frac{1}{2} \ln \theta - \ln \gamma + \ln \rho \big ). $

\section{4. An application: quantum oscillations in the ultrasound velocity}

To make the presented results more specific  we apply  our analysis to study the
effects of the FS local geometry on the quantum oscillations 
of the velocity of ultrasound waves travelling in a metal. 
When an ultrasound wave propagates in a metal the crystalline 
lattice is periodically deformed. It gives rise to electric  
fields which influence the electrons. Besides, the 
periodical deformations of the lattice cause changes in the 
electronic spectrum. Here, we omit these deformation 
corrections to simplify further analysis.

The emergence of the electric field accompanying the lattice deformation,       
leads to a redistribution of the electron density $ N .$ The
local change in the electronic density $ \delta N \bf (r) $
 equals:
 % f17,19,17
      \be
\delta N ({\bf r}) =  \frac{\partial N}{\partial \zeta}
e \varphi ({\bf  r}) 
 \equiv  N_\zeta
e \varphi ({\bf r}) . 
                               \ee
  The relation (17) has to be complemented by the 
condition of electrical neutrality of the system:
      %f18,20,18
   \be
\delta N ({\bf r}) -
e N {\mbox {div}} {\bf  u} ({\bf  r}) = 0 .
           \ee
 
 We use these equations to express the 
potential $ \varphi ({\bf  r}) $ in terms of the lattice 
displacement vector. As a result we  derive the expression 
for the electron force {\bf F}({\bf r}) acting upon the 
lattice under its displacement by the vector 
{\bf u}({\bf r}). For a longitudinal ultrasound wave 
travelling along the magnetic field $ \bf B $ we obtain:
  \be % f19,21,19
 F {\bf (r)} = - \frac{N^2}{N_\zeta} {\bf b}_{0} 
\Delta \bf  u (r) .
  \ee
   Here, $ \bf b_0 $ is unit vector directed along the 
field $ \bf B.$ More thorough analysis taking into account 
deformation corrections to the electron energies does not 
bring qualitative changes in the expression (19). 
We arrive at the modified expression for the force 
$ F \bf (r)$ simply multiplying (19) by a dimensionless 
factor of the order of unity which could be treated as a 
constant. 

Starting from (19) we obtain the well known result for the 
oscillating correction $ \tilde s $ to the sound velocity 
$ s:$
   \be % f20,22,1
  \frac{\tilde s}{s} = - \frac{N^2}{\rho s^2 g} 
\sum_i \Delta_i
  \ee
     where
   $ \rho $ is the metal density and
 the summation is carried out over all extremal 
cross-sections on the FS corresponding to the given direction 
of the magnetic field. So, the above angular dependences of 
magnitudes of quantum oscillations arising due to the local  
geometry of the FS could be revealed in oscillations of the 
velocity of ultrasound waves propagating in metals.

\section{5. Conclusion}

In summary, we showed that when the FS of a metal inserts lines where its curvature turns zero or reveals discontinuity, this can significantly affect both amplitude and phase of electron DOS oscillations in strong (quantizing) magnetic fields. The effect arises due to increase/decrease of the relative number of electrons associated with the neighborhoods of these lines. It reveals itself at certain directions of the magnetic field, and disappears when the magnetic field is tilted away from such direction. This results in angular dependence of the DOS oscillation amplitudes. The latter could be observed in experiments on quantum oscillations in various characteristics of a metal such as magnetization, magnetic susceptibility and resisitivity, for all these oscillations arise due to the DOS quantum oscillations. Quantum oscillations in the sound velocity discussed in this work is a mere example of the effect of the FS curvature local anomalies on observables.

The most obvious
materials where we can expect the FS curvature anomalies 
to be manifested include layered 
structures with metallic-type conductivity 
(e.g. $\alpha-(BEDT-TTF)_2MHg(SCN)_4 $ group of organic 
metals). Fermi surfaces of these materials are sets of 
rippled cylinders, isolated or connected by links 
\cite{20}.  The angular dependence of the magnetic oscillations 
magnitude resembling that shown in the Fig. 3 is well known 
for such materials. The appearance of this effect shows that 
the quasi-two-dimensional FSs of some organic metals include 
segments with zero curvature. The effect was first analyzed 
in \cite{21}, and further discussed in some later works (see e.g. \cite{20}). 

Local 
geometry of the FSs of usual metals also can be displayed 
in the angular dependences of magnitudes of quantum 
oscllations.
For instance, there is an 
experimental evidence that ``necks" connecting quasispherical 
pieces of the FS of copper include nearly cylindrical belts 
\cite{1}. When the magnetic field is directed along the 
axis of  a ``neck" (for instance, along the [111] direction in the quasimomenta space), the extremal cross section of the 
``neck" could be expected to run along the nearly cylindrical 
strip where the FS curvature turns zero. It is also likely 
that the FS of gold possesses the same geometrical features 
for it closely resembles that of copper. As for possible 
divergences of the FS curvature, experiments 
of \cite{18} give grounds to conjecture that such 
anomalies could be found on the FSs of cadmium and zink. It 
is possible that the curvature is anomalously large at the 
edge of the electron lens which is the part of the FSs in 
both metals. For all above listed 
substances we can expect the effect to be revealed at 
reasonably low temperatures $ (\sim 1K)$ and reasonably 
strong magnetic fields $ (1 \div 10 T),$  giving additional information on the FSs fine geometrical characteristics.

\section{Acknowledgments}
I thank G.M. Zimbovsky for help with the manuscript. This work was supported  in part by NSF Advance  program SBE-0123654 and PR Space Grant NGTS/40091.

\end{document}